\documentclass[fleqn,10pt]{wlscirep}
\usepackage[utf8]{inputenc}
\usepackage[T1]{fontenc}
\usepackage{graphicx}
\usepackage[utf8]{inputenc}
\usepackage{algpseudocode}
\usepackage[ruled,vlined]{algorithm2e}
\usepackage{float}
\usepackage{svg}
\usepackage{cite}
\usepackage{hyperref}
\usepackage{comment}
\usepackage{multirow}
\usepackage{caption} 

\title{Enhanced ECG Arrhythmia Detection Accuracy by Optimizing Divergence-Based Data Fusion}
\author[1]{Baozhuo Su}
\author[2]{Qingli Dou}
\author[1,a]{Kang Liu}
\author[1]{Zhengxian Qu}
\author[1]{Jerry Deng}
\author[1,b]{Ting Tan}
\author[2,c]{Yanan Gu}

\affil[1]{Aerie Intelligent Technology Corp, Irvine, CA, USA}
\affil[2]{Department of Emergency Medicine, People’s Hospital of Shenzhen Baoan District Shenzhen, The Second Affiliated Hospital of Shenzhen University, Shenzhen,  Guangdong, China}

\affil[a]{Corresponding author. Email: kliu@aerieintech.com}
\affil[b]{Corresponding author. Email: tting@aerieintech.com}
\affil[c]{Corresponding author. Email: gyn1252006@163.com}

\begin{abstract}
AI computation in healthcare faces significant challenges when clinical datasets are limited and heterogeneous. Integrating datasets from multiple sources and different equipments is critical for effective AI computation but is complicated by their diversity, complexity, and lack of representativeness, so we often need to join multiple datasets for analysis. The currently used method is fusion after normalization. But when using this method, it can introduce redundant information, decreasing the signal-to-noise ratio and reducing classification accuracy. To tackle this issue, we propose a feature-based fusion algorithm utilizing Kernel Density Estimation (KDE) and Kullback-Leibler (KL) divergence. Our approach involves initially preprocessing and continuous estimation on the extracted features, followed by employing the gradient descent method to identify the optimal linear parameters that minimize the KL divergence between the feature distributions. Using our in-house datasets consisting of ECG signals collected from 2000 healthy and 2000 diseased individuals by different equipments and verifying our method by using the publicly available PTB-XL dataset which contains 21,837 ECG recordings from 18,885 patients. We employ a Light Gradient Boosting Machine (LGBM) model to do the binary classification. The results demonstrate that the proposed fusion method significantly enhances feature-based classification accuracy for abnormal ECG cases in the merged datasets, compared to the normalization method. This data fusion strategy provides a new approach to process heterogeneous datasets for the optimal AI computation results.\\
\\

\end{abstract}
\begin{document}
\flushbottom
\thispagestyle{empty}
\maketitle
\section*{Introduction}
Algorithmic bias in AI computation often stems from the lack of diversity and representativeness in datasets. When AI computation is used in the field of healthcare, challenges are particularly posed with heterogeneous small clinical datasets, as large volumes of high-quality data are required to train effective models in the context of machine learning. Thus, to integrate clinical datasets from various sources for efficient AI computation is not only critical but also challenging due to the heterogeneity, complexity, and insufficient data representativeness of most clinical datasets\cite{0}. To address this type of issues involved in medical AI computation, a specialized data quality framework, the METRIC-framework was recently proposed by Schwabe and colleagues~\cite{npj}. Through evaluating the diversity of datasets across demographic factors, data sources, and accuracy of the measurements relevant to both device and human errors and so forth, the METRIC framework offers a systematic approach to evaluating data quality and reducing biases of small clinical datasets without providing standardized guidelines for the measurement and assessment of data quality. This issue was resolved by using a standardized dataset based on the Fast Healthcare Interoperability Resources (FHIR) standard, which improves data interoperability and sharing across various systems~\cite{FHIR}. AI computation has been tested to analyze electrocardiograph (ECG) data for the detection and classification of arrhythmias in clinical settings ~\cite{8,9}. The methods used in this application involved either feature-based approaches, which combined manually crafted features with traditional machine learning algorithms, or deep learning techniques for automatic extraction of ECG features. However, these methods were effective only for a single dataset~\cite{12}. A hybrid method combining deep learning with data was recently tested for ECG arrhythmia analysis~\cite{17}. While this method offered certain advantages by integrating two lightweight deep learning models into a unified framework, it still depended on a single dataset and did not incorporate data fusion across multiple datasets. To overcome the obstacle of data fusion across multiple datasets for AI computation training, several different approaches have been recently investigated in clinical ECG datasets. These efforts include using deep learning models and machine learning algorithms to process multi-sourced datasets~\cite{18,19}. These studies were primarily focused on the role of selected factors in data fusion or the methods to train models for better classification outcomes, but they often overlooked the negative impact of information redundancy and errors from multi-source datasets, which could lead to inaccurate outcomes when merged datasets were re-analyzed. To address these issues, we propose a novel method that combines Kullback-Leibler (KL) divergence optimization and machine learning techniques for fusing ECG feature datasets. We first measure the distribution discrepancy between two datasets utilizing the KL divergence. Specifically, a Gaussian kernel-based estimation of distributions is conducted to identify and minimize the inter-dataset parameters with a linear relationship to achieve a minimal KL divergence. This innovative strategy is designed to leverage the strengths of both traditional and modern machine learning techniques to enhance the classification accuracy and robustness of arrhythmia detection systems by effectively reducing the redundancy and errors present in multi-source datasets.

\begin{figure}[htbp]
\centering
\includegraphics[width=1.1\textwidth]{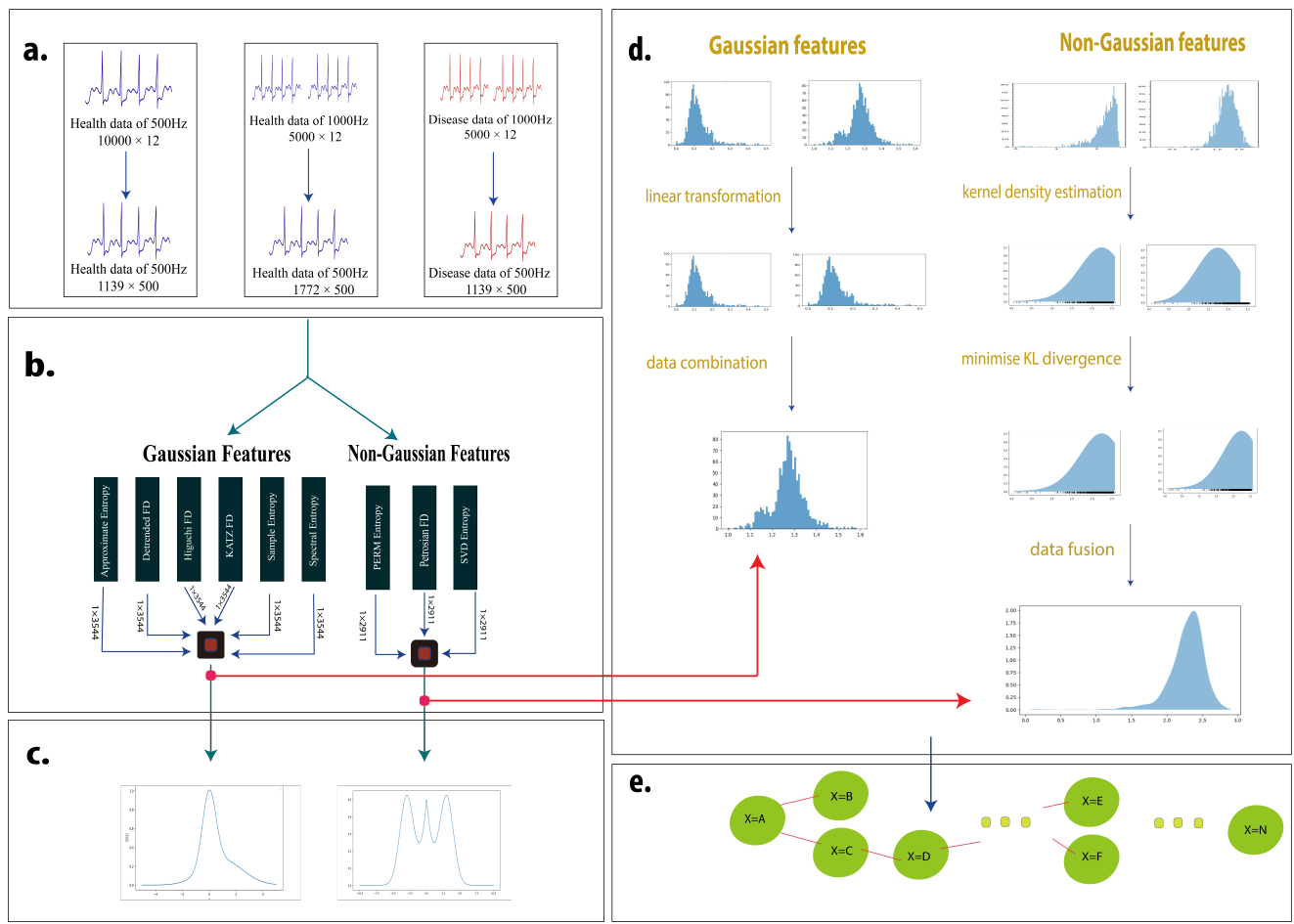}
  \label{fig:Fig 1}
   \caption{Schematic of the overall workflow: \textbf{(a)} Initial normalization to covert all 12-lead ECG recordings to 500 Hz. \textbf{(b)} Extraction of six Gaussian and three non-Gaussian features. \textbf{(c)} categorized into Gaussian and Non-Gaussian features. \textbf{(d)} Gaussian features are applied the  rigid linear transformation, while Non-Gaussian features are processed using KDE followed by divergence optimization. \textbf{(e)} All features processed were fed into LGBM to train the model.}
\end{figure} 

\section*{RESULTS}
\vspace{-0pt} 
\subsection*{Optimized data fusion algorithms increase the accuracy of classification in in-house datasets}
For features that adhere to a Gaussian distribution, we leverage the property that Gaussian distributions retain their natures when subjected to an linear transformation. Conversely, for features exhibiting non-Gaussian distributions, we endeavored to minimize the KL divergence between the same feature's distribution of different datasets. We will discuss how the features are selected in the section Datasets and Methods.

Upon successfully fusing the health datasets, we leverage the optimized features from the combined healthy dataset and contrast them with features from the unhealthy dataset to train the LGBM model. This model is then tasked with executing classification tasks, aiming to distinguish between healthy and unhealthy data instances based on the derived features.

Table~\ref{table:table1} lists the classification accuracies of the LGBM model for each extracted feature. In addition, it compares these results with the classification results of the dataset that was subjected to conventional standardization before merging. 

Each model went through 100 training iterations. The table below lists the average accuracies derived from these 100 training sessions, which together reflect the performance of the models over multiple training sessions. This approach reduces the impact of any potential outliers or anomalies in the dataset, thus ensuring the robustness and reliability of the findings.
For internal data, the highest classification accuracy achieved is 99\%, which is observed with the Permutation Entropy and Approximate Entropy feature. This indicates a significant improvement in the accuracy of classification when data fusion techniques are applied. Other notable feature is the Singular Value Decomposition Entropy, which achieves the Data Fusion Accuracies of 86\%.

On the other hand, the standardized classification accuracy values of the raw data do not vary over a wide range. Spectral Entropy has the highest normalized accuracy of 69\%,  while all other features have a lower classification accuracy. These variations suggest that the features themselves are not better discriminators, but the application of our data fusion technique greatly improves their effectiveness.

\begin{figure}[htbp]
  \includegraphics[width=1\textwidth]{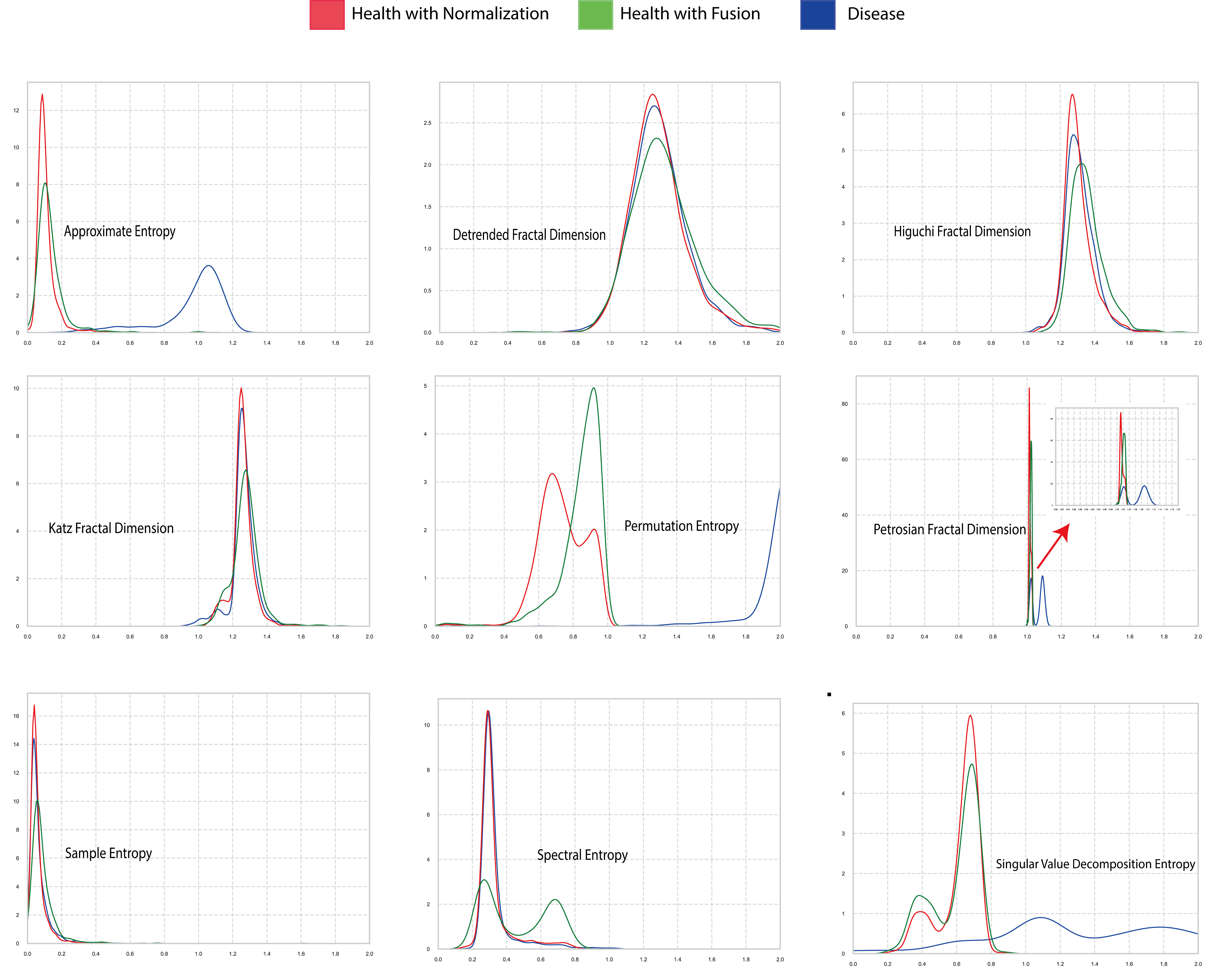}
   \caption{Feature Distributions of in-house datasets.
     The distribution of features are presented after the KDE estimation. This figure illustrates the density distribution of different feature extracted for In-house datasets.}
  \label{fig:Fig 2}
\end{figure}

\begin{figure}[htbp]
  \centering
  \includegraphics[width=0.75\textwidth]{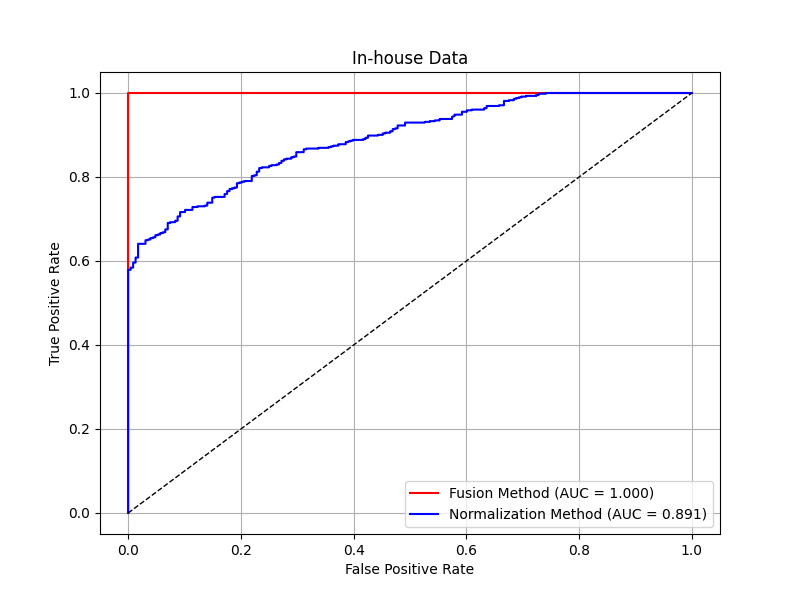}
   \caption{The prediction performance for In-house datasets.
  The performance of LGBM model for In-houe datasets is illustrated by the ROC cureve, comparing the nomalization method and divergence-based optimazation method. }
  \label{fig:Fig 3}
\end{figure}

\begin{table}[htbp]
\vspace{10pt} 
\centering
\caption{Detailed feature-based classification of in-house datasets}
\normalsize
\vspace{10pt} 
\begin{tabular}{|c|c|cc|cc|}
\hline
\textbf{Distribution} & \textbf{Feature} & \multicolumn{2}{c|}{\textbf{Normalization}} & \multicolumn{2}{c|}{\textbf{Data Fusion}} \\
\cline{3-6}
 &  & Accuracy (\%) & FPR/FNR (\%) & Accuracy (\%) & FPR/FNR (\%) \\
\hline
\multirow{6}{*}{\textbf{Gaussian}} & \textbf{Approximate Entropy} & 34 & 71.52/61.48 &\textbf{99} & \textbf{0.80/0.01} \\
& Detrended Fractal Dimension & 33 & 68.84/64.9 & 56 & 39.61/48.12 \\
& Higuchi Fractal Dimension & 35 & 62.22/68.46 & 71 & 28.53/28.53 \\
& Katz Fractal Dimension & 35 & 71.11/59.69 & 67 & 29.32/35.92 \\
& Sample Entropy & 35 & 70.62/59.71 & 75 & 24.55/24.55 \\
& Spectral Entropy & 69 & 33.31/28.52 & 79 & 16.72/26.35 \\
\hline
\multirow{3}{*}{\textbf{Non-Gaussian}} & Petrosian Fractal Dimension & 36 & 61.42/66.23 & 36 & 64.23/63.44 \\
& \textbf{Permutation Entropy} & 34 & 68.42/63.25 & \textbf{99} & \textbf{0.76/0.01} \\
& Singular Value Decomposition Entropy  & 32 & 68.82/67.93 & 86 & 21.91/7.12 \\
\hline
\multirow{1}{*}{\textbf{Mixed}} & Combined & 63 & 36.42/37.74 & 100 & 0/0 \\
\hline
\end{tabular}
\label{table:table1}
\\
\parbox{\linewidth}{\footnotesize *\textbf{FPR}: False positive rate}
\parbox{\linewidth}{\footnotesize *\textbf{FNR}: False negative rate}
\end{table}

\subsection*{Verification Experiment by using PTB-XL Dataset}
To validate the effectiveness and generalizability of our newly proposed ECG signal processing algorithms, it is essential to test and verify them on different datasets. Choosing an appropriate dataset is a crucial aspect of experimental design. In this study, we selected the PTB-XL dataset for our experiments.
The PTB-XL dataset, released by the Philipps-Universität Marburg and Technische Universität Berlin, is one of the largest publicly available ECG datasets. It contains 21,837 12-lead ECG records from 18,885 patients, each record lasting 10 seconds. The dataset not only includes raw ECG signals, but also provides detailed diagnostic labels covering various cardiac conditions such as myocardial infarction.
In this work, we utilized the healthy data from the PTB-XL dataset and integrated it with clinical healthy datasets, while performing a binary classification experiment in-house disease datasets. It is important to note that the PTB-XL dataset consists of data collected from various instruments, but this factor was not considered in this study. Also noting that, as the in-house data do not have the subdivided classification of arrythmia, so that here we do not distinguish different types of arrythmia either, although the PTB-XL dataset gives a very detailed classification.
We used the same process for the PTB-XL dataset and obtain the results shown in Table~\ref{table:table2}:

\begin{figure}[htbp]
  \centering
  \includegraphics[width=1\textwidth]{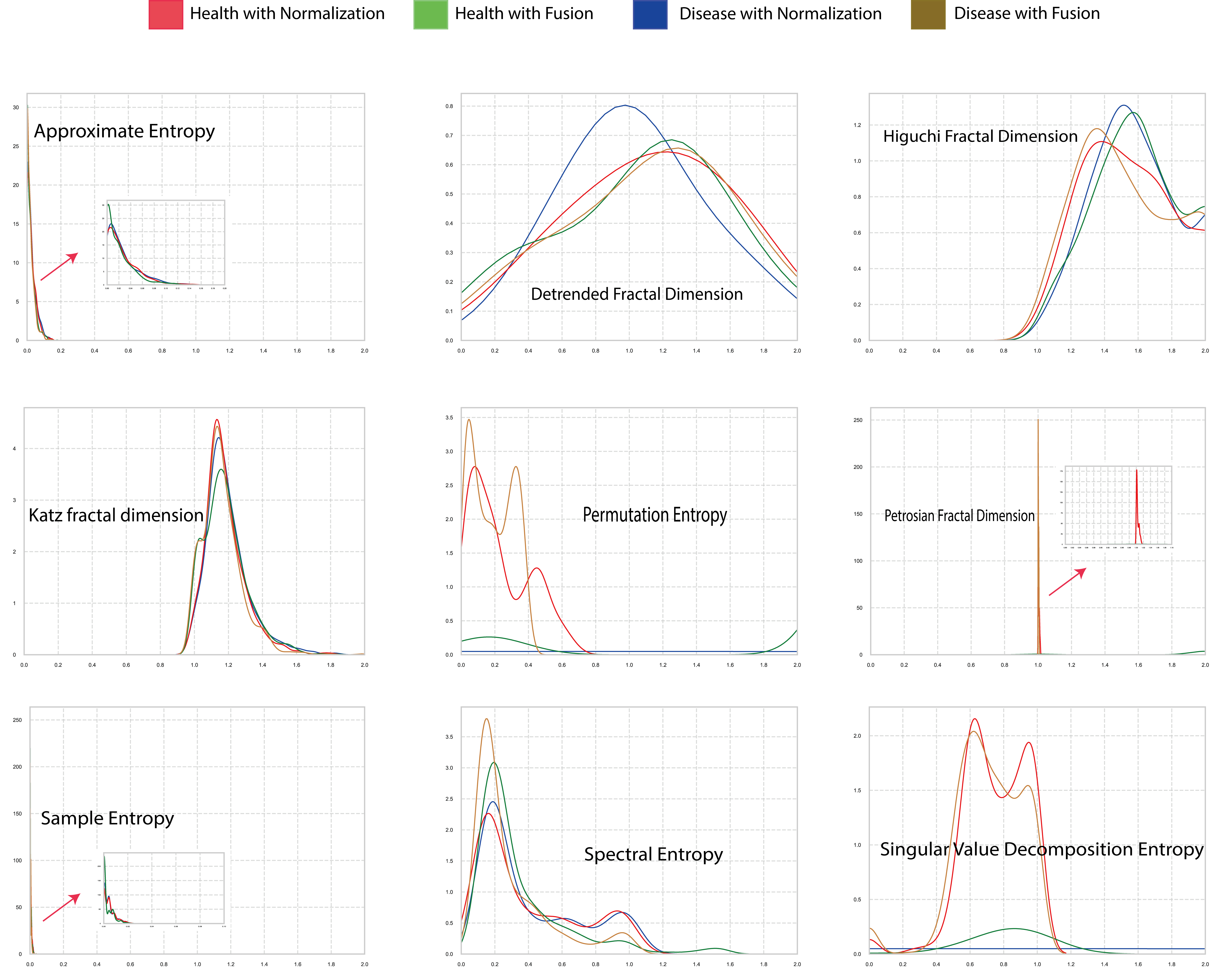}
   \caption{ Feature Distributions of PTB-XL datasets.
     The distribution of features are presented after the KDE estimation. This figure illustrates the density distribution of different feature extracted for PTB-XL datasets.}
  \label{fig:Fig4}
\end{figure}

\begin{figure}[htbp]
  \centering
  \includegraphics[width=0.75\textwidth]{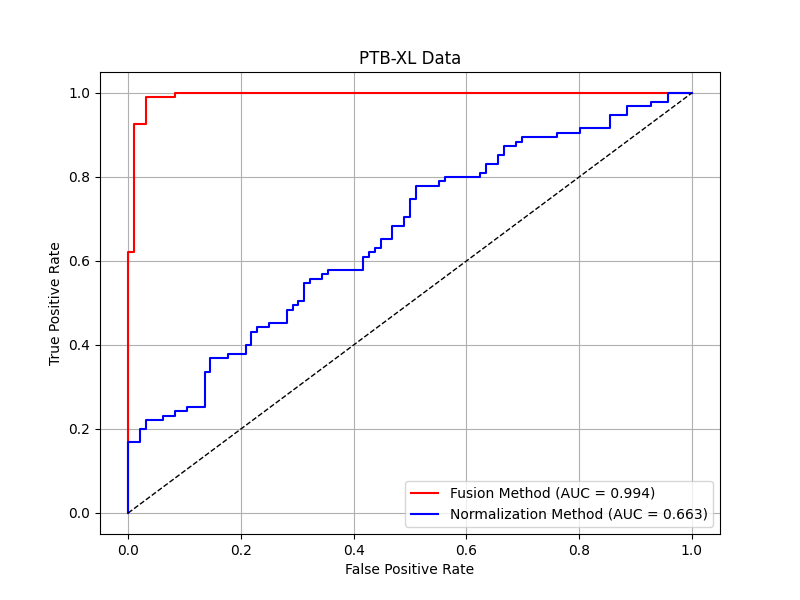}
  \caption{The prediction performance for PTB-XL datasets. The performance of the LGBM model for PTB-XL datasets is illustrated by the ROC curve, comparing the normalization method and divergence-based optimization method, which includes two datasets for each method, as shown in the appendix.} 
  \label{fig:Fig 5}
\end{figure}

\begin{table}[htbp]
\vspace{10pt} 
\centering
\caption{Detailed feature-based classification of PTB-XL datasets}
\normalsize
\vspace{10pt} 
{
\begin{tabular}{|c|c|cc|cc|}
\hline
\textbf{Distribution} & \textbf{Feature} & \multicolumn{2}{c|}{\textbf{Normalization}} & \multicolumn{2}{c|}{\textbf{Data Fusion}} \\
\cline{3-6}
 &  & Accuracy (\%) & FPR/FNR (\%) & Accuracy (\%) & FPR/FNR (\%) \\
\hline
\multirow{6}{*}{\textbf{Gaussian}} & Approximate Entropy & 52 & 46.31/49.40 & 65 & 35.17/35.54\\
& Detrended Fractal Dimension & 60 & 50.52/33.92 & 60 & 47.58/37.97 \\
& Higuchi Fractal Dimension & 52 & 56.84/42.85 & 53 & 51.03/44.59 \\
& \underline {Katz Fractal Dimension} & 57 & 43.15/43.45 & \underline{56} & \underline{43.44/43.55} \\
& Sample Entropy & 60 & 44.21/37.50 & 79 & 17.93/22.99 \\
& Spectral Entropy & 57 & 46.31/40.40 & 66 & 39.31/31.35 \\
\hline
\multirow{3}{*}{\textbf{Non-Gaussian}} & \textbf{Petrosian Fractal Dimension} & 68 & 40.00/27.97 & \textbf{91} & \textbf{1.05/13.69} \\
& \textbf{Permutation Entropy} & 63 & 34.73/37.50 & \textbf{90} & \textbf{6.31/1.25} \\
& \textbf{Singular Value Decomposition Entropy} & 52 & 52.63/44.64 & \textbf{90} & \textbf{7.36/10.71} \\
\hline
\multirow{1}{*}{\textbf{Mixed}} & Combined & 75 & 36.84/17.85 & 95 & 1.05/7.73 \\
\hline
\end{tabular}
\label{table:table2}
}
\end{table}

In the PTB-XL dataset, the Spectral Entropy and Permutation Entropy feature achieve the highest Data Fusion Accuracy of 99\%, reflecting the efficiency of these features in capturing relevant patterns for accurate classification. 
\subsection*{Fusion between PTB-XL and In-house datasets}
In order to make our experience more complete, we have used one health data and one unhealthy data from each of the internal datasets and PTB-XL datasets, and finally comparing the results of 3 experiments. The results are shown in Table~\ref{table: In-house and PTBXL}. In conclusion, we can derive that although some of the features do not show the better performance after processing, this may due to the raw attributes of PTB-XL dataset, but in a clinical situation, we will consider to use combined feature which has a huge improvement after processing in the three cases above.

\begin{figure}[htbp]
  \centering
  \includegraphics[width=1\textwidth]{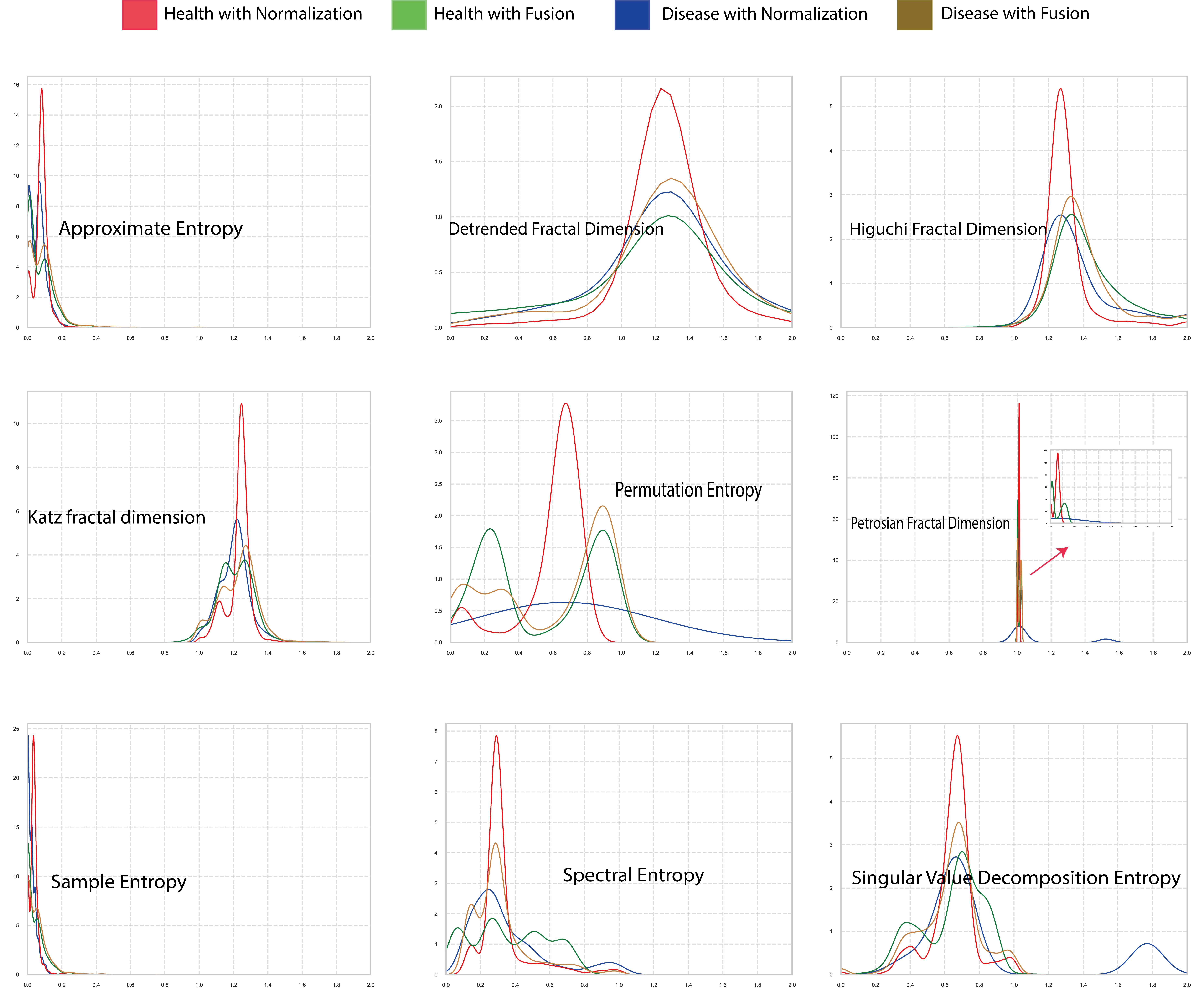}
   \caption{Feature Distributions of both using In-house and PTB-XL datasets.
     The distribution of features are presented after the KDE estimation. This figure illustrates the density distribution of different feature extracted.}
  \label{fig:Fig6}
\end{figure}

\begin{figure}[htbp]
  \centering\includegraphics[width=0.75\textwidth]{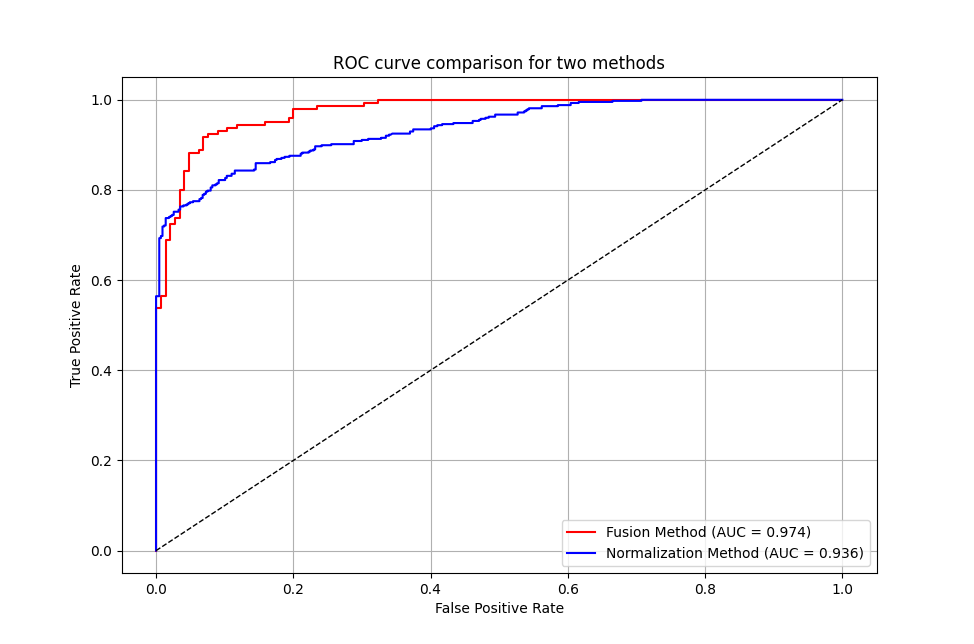}
  \caption{The prediction performance for both using In-house and PTB-XL datasets. 
  The performance of LGBM model when both using In-house and PTB-XL dataets is illustrated by the ROC cureve, comparing the nomalization method and divergence-based optimazation method, where including 2 datasets as showed in appendix for each methods.}
  \label{fig:Fig 7}
\end{figure}

\begin{table}[htbp]
\vspace{10pt} 
\centering

\caption{Detailed feature-based classification of fusion between PTB-XL and in-house datasets}
\normalsize
\vspace{10pt} 

\begin{tabular}{|c|c|cc|cc|}
\hline
\textbf{Distribution} & \textbf{Feature} & 
\multicolumn{2}{c|}{\textbf{Normalization}} &
\multicolumn{2}{c|}{\textbf{Data Fusion}}
\\
\cline{3-6}
 &  & Accuracy (\%) & FPR/FNR (\%) & Accuracy (\%) & FPR/FNR (\%) \\
\hline
\multirow{6}{*}{\textbf{Gaussian}} & Approximate Entropy & 69 & 31.38/31.14 & 70 & 31.72/29.81 \\

& \underline{Detrended Fractal Dimension} & 59 & 42.85/38.87 & \underline{56} & \underline{45.51/44.06} \\

& \underline{Higuchi Fractal Dimension} & 72 & 30.91/25.76 & \underline{58} & \underline{48.96/40.53} \\

& \underline{Katz Fractal Dimension} & 66 & 31.85/35.59 & \underline{56} & \underline{42.75/40.11} \\

& Sample Entropy & 74 & 22.48/31.61 & 80 & 27.59/18.07 \\

& Spectral Entropy & 63 & 37.93/36.29 & 74 & 39.31/23.58 \\
\hline
\multirow{3}{*}{\textbf{Non-Gaussian}} & \textbf{Petrosian Fractal Dimension} & 79 & 24.82/17.09 & \textbf{94} & \textbf{1.87/9.79} \\

& \textbf{Permutation Entropy}  & 78 & 24.12/19.91 & \textbf{99} & \textbf{0.01/0.00} \\

& Singular Value Decomposition Entropy & 61 & 46.37/31.61 & 79 & 6.32/31.44 \\
\hline
\multirow{1}{*}{\textbf{Mixed}} & Combined & 86 & 11.71/16.86 & 95 & 6.21/4.98 \\
\hline
\end{tabular}

\label{table: In-house and PTBXL}
\end{table}

\section*{DISCUSSION}
By integrating clinical ECG datasets from multiple sources and optimizing feature distributions, we demonstrate that divergence-based data fusion can significantly improve ECG classification performance. Our approach leverages the strengths of both Gaussian and non-Gaussian feature transformations, ensuring minimal redundancy while preserving essential information for machine learning models. Specifically, features such as Permutation Entropy and Approximate Entropy exhibit substantial gains in accuracy after data fusion, achieving nearly perfect classification in in-house datasets. The integration of PTB-XL data further confirms our approach, as the results demonstrate notable improvements over conventional normalization-based fusion strategies. These findings suggest that our divergence-optimized fusion approach effectively mitigates the inconsistencies introduced by heterogeneous datasets. This study highlights the significance of data fusion techniques in enhancing ECG arrhythmia detection accuracy.

Despite these promising outcomes, certain limitations must be acknowledged. While selected features exhibit significant improvements in classification performance, others do not respond as favorably to the fusion process. For instance, the Singular Value Decomposition Entropy performs well in in-house datasets but does not yield similar benefits in the PTB-XL datasets. This discrepancy suggests that underlying dataset characteristics, including variations in data acquisition and preprocessing, may influence the effectiveness of specific features. Another key observation is that the magnitude of classification improvement correlates with the extent of feature distribution alignment before and after fusion. Furthermore, we have observed that the optimized method have shown a better performance on features with non-Gaussian distribution than those with Gaussian distribution, the explanation may be one of our future work. Features that exhibit substantial shifts in their distributions tend to show greater gains in classification accuracy, supporting the hypothesis that optimizing distributional alignment is critical for effective data fusion. Further investigation is needed to quantify these relationships more precisely and develop feature-specific optimization strategies. Future work may also focus on refining our method through automatic selection of transformation parameters and hyperparameter tuning to improve convergence. Additionally, extending this approach to other biomedical signals, such as EEG, may further validate its applicability in broader clinical contexts. Exploring deep learning-based fusion techniques could also complement our current approach, offering more adaptive methods for handling heterogeneous datasets.

In conclusion, our findings establish a solid foundation for divergence-based data fusion in medical AI applications. By addressing the challenges of multi-source data integration, this study contributes to the advancement of robust and reliable machine learning models for clinical decision-making. With continued development, these techniques hold the potential to improve diagnostic accuracy and facilitate personalized healthcare solutions.

\vspace{-0mm} 
\section*{Datasets and Methods}
\vspace{-0mm} 
\subsection*{Construction of in-house datasets}
The current study first tests the optimized LGBM model in internal datasets comprising ECG recordings from two healthy cohorts and one arrhythmia cohort. Each healthy cohort includes 1,000 ECG recordings, while the arrhythmia cohort consists of 2,000 clinical ECG recordings. The recordings of a healthy cohort were captured using a 12-lead ECG machine at a sampling rate of 500 Hz (Nihon Kohden ECG-2350, Tokyo, Japan), and the other healthy cohort and arrhythmia recordings were collected at a sampling rate of 1,000 Hz (Mindray BeneHeart R12A, Mindray Bio-Medical Electronics, Shenzhen, China). The arrhythmia cohort includes multiple forms of dysrhythmic conditions such as ectopic beats, tachycardia, bradycardia, sinus arrest, premature beats, and atrial fibrillation. Each recording is a 10-second 12-lead ECG. A total of 4,000 ECG recordings, representing 4,000 individual subjects (one recording per subject), were randomly selected from a large pool of 17,363 recordings collected by the Department of Emergency Medicine at Shenzhen Baoan People’s Hospital during 2022 and 2023. The initial diagnoses were made by the software of the ECG machine and confirmed by two experienced ECG experts. The study was approved by the Ethics Committee, the institutional review board (IRB) of Shenzhen Baoan People’s Hospital, and all participants signed an informed consent statement before data collection.

\subsection*{PTB-XL datasets and data extraction for the study}
The PTB-XL datasets are a large publicly available ECG collection that contains 21,837 clearly annotated 10-second length 12-lead ECG waveform recordings from 18,885 patients and multiple different types of ECG machines. In addition to the recordings from healthy subjects, the datasets include arrhythmias and various other cardiac conditions such as myocardial infarction. In the current, 2000 healthy and 2000 arrhythmia recordings are randomly extracted from the PTB-XL datasets to match the size of the in-house datasets.  
\subsection*{Data processing}
All ECG recording data are subject to a sequential process of sampling rate normalization, downscaling (reduction of dimensionality) coupled with resampling, denoising, and slicing of ECG signals before selection and extraction of features for analysis. Briefly, to simplify computational demands and streamline the processing framework, the sampling rates of all ECG recordings are adjusted from 1,000 Hz to 500 Hz for both in-house and extracted PTB-XL datasets using downsampling methods followed by denoising by Fast Fourier Transform (FFT). In this regard, a defined threshold is used to identify and eliminate noisy elements in the data. To precisely detect the rate variability of heartbeats, an appropriate window size and sampling frequency are used to mark the regions of interest within each ECG signal to isolate heartbeat peaks. This adjustment ensures that each processed ECG signal extends uniformly to 500 samples and thus establishes a consistent baseline for further examination.
\subsection*{Feature selection and extraction}
A total of nine specific ECG-related features (Table~\ref{table:ecg_features}) are selected for extraction. These features cover a wide spectrum of cardiac activity and function. Each of these features has been previously validated in clinical research for their diagnostic value in various cardiac conditions and is extracted from the refined datasets using the python library entropy for subsequent model training and analysis. 

\begin{table}[htbp]
\centering
\caption{Summary of Selected ECG-related Features.}This table provides a concise overview of key ECG signal features, with each description summarized by two to three detailed keywords. These keywords highlight specific aspects of each feature, such as types of complexity or specific changes in the signal, offering a clear reference for their role in ECG analysis.
\scalebox{0.8}{
\begin{tabular}{|c|p{13.5cm}|}
\hline
\textbf{Feature Name} & \textbf{Feature Description } \\ \hline
Approximate entropy\cite{APP} & A statistical measure used to quantify the regularity and complexity of time series ECG data and is effective for the detect of arrhythmias  \\ \hline
Detrended fractal dimension\cite{DETREND} & A measure used to analyze the complexity and self-similarity of a time series ECG signal after removing trends \\ \hline
Higuchi fractal dimension\cite{higuchi} & A measure used to quantify the complexity of a time series ECG signal \\ \hline
Katz fractal dimension\cite{KATZ} & A measure used to estimate the fractal dimension of a time series ECG signal \\ \hline
Petrosian fractal dimension\cite{PETRO} & A measure used to estimate the fractal dimension of a time series ECG signal helping to distinguish between different types of signals based on their complexity and self-similarity \\ \hline 
Permutation entropy\cite{PERM} & Permutation-based complexity, order of signal values \\ \hline
Sample Entropy\cite{SAMPLE} & A statistical measure used to quantify the complexity and irregularity of time series is particularly effective for detecting irregular heart rhythms on ECG \\ \hline
Spectral Entropy\cite{SPEC} & A measure of the complexity or randomness of a signal derived from its power spectrum, quantifying the distribution of power across different frequency components of ECG signals \\ \hline
Singular value decomposition entropy\cite{SVD} & A measure of the complexity and information content of a time series ECG derived from Singular Value Decomposition (SVD) helps to understand the structure of the ECG signal and detect anomalies \\ \hline
\end{tabular}}
\label{table:ecg_features}
\end{table}

\begin{figure}[htbp]
  \includegraphics[width=\textwidth]{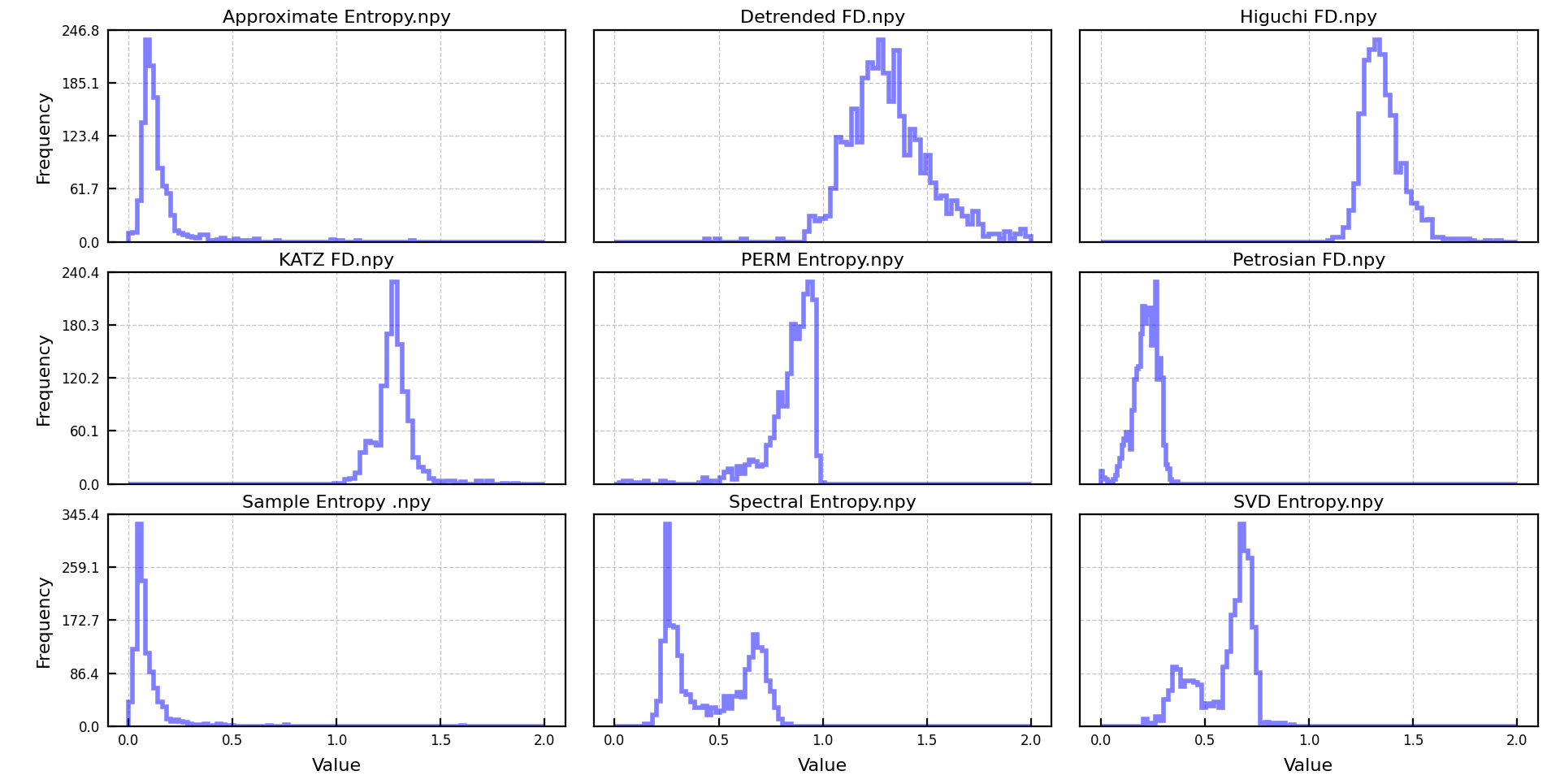}
  \caption{The figure illustrates nine features extracted from the unhealthy dataset after preprocessing. It shows that Approximate Entropy, Detrended Fractal Dimension, Higuchi Fractal Dimension, Katz Fractal Dimension, Sample Entropy, and Spectral Entropy exhibit Gaussian or Multi-Gaussian distributions. In contrast, Petrosian Fractal Dimension, Permutation Entropy, and Singular Value Decomposition Entropy display non-Gaussian distributions.}
  \label{fig:Fig8}
\end{figure}

Feature extraction is the process of identifying and extracting useful information from raw data for subsequent analysis and model training. We extracted nine entropy features from the three preprocessed datasets described above using the python library 'antropy' \cite{antropy}. Figure~\ref{fig:Fig8} has shown the distribution of each feature extracted from unhealthy data being processed.

\subsection*{Data Fusiuon}
Data fusion is processed by optimized LGBM modeling. We adopt a bifurcated strategy for the data fusion algorithm, treating Gaussian and non-Gaussian distributed features distinctly to optimize the fusion process for each case. 
Consider two sets of the same features, denoted as $X$ and $Y$ respectively. The primary objective of the data fusion method is to determine an affine transformation that maps set $Y$ to a new set $Y'$ in such a way that the distributions of elements in $X$ and $Y'$ become similar. In other words, we aim to find suitable values for $C$ and $D$ to satisfy the following equation:
\begin{equation}\label{eq:main}
\text{Law}_{y\in Y} (Cy+D) \approx \text{Law}_{x\in X}(x).
\end{equation}
As mentioned in the previous section, we adopt a bifurcated strategy for the data fusion algorithm, treating Gaussian and non-Gaussian distributed features distinctly to optimize the fusion process for each case.

\subsubsection*{Situation A: Features with Gaussian Distribution}
For Gaussian distributed features, we leverage the property that Gaussian distributions retain their natures when subjected to an affine transformation. The explicit formula for the optimal transformation parameters $C^{*}$ and $D^{*}$ is as follows:
\begin{equation}\label{eq:gauss}
    C^{*} = \sigma_X^{1/2} \sigma_Y^{-1/2}, \qquad D^{*} = \mu_X - \sigma_X^{1/2} \sigma_Y^{-{1/2}} \mu_Y,
\end{equation}
where $\mu_X$ (resp.\@ $\mu_Y$) and $\sigma_X$ are the expectation and the covariance matrix of $X$ (resp.\@ $Y$).

\subsubsection*{Situation B: Features with non-Gaussian distribution}
When dealing with features displaying non-Gaussian distributions, we encounter two primary challenges in the context of equation \eqref{eq:main}. The first challenge lies in the difficulty of quantifying these distributions, and the second revolves around interpreting the nature of the approximation in equation \eqref{eq:main}. 
These two issues will be addressed in the following two steps:
\begin{enumerate}
    \item Employ Kernel Density Estimation (KDE) method to estimate the continuous distribution of elements in both sets $X$ and $Y$. For instance, in the case of $N$ data points, denoted as $x_1, \ldots, x_N$, the KDE-estimated density $\hat{f}$ is given by:
    \begin{equation*}
        \hat{f}(x) = \frac{1}{N}\sum_{i=1}^{N} \frac{1}{\sqrt{2\pi}\sigma} \exp\left(-\frac{\|x - x_i\|^2}{2\sigma^2}\right),
    \end{equation*}
  where $\sigma$ is the bandwidth.
    \item Employ the the Kullback-Leibler (KL) divergence to quantify the ``distance" of two estimated distributions. Given two probability density functions \( f \) and \( g \) defined in $\mathbb{R}^d$, the KL divergence from \( f \) to \( g \) is given by:
\[
D_{\text{KL}}(f || g) = \int_{\mathbb{R}^d} f(x) \log\left(\frac{f(x)}{g(x)}\right) dx.
\]
    \end{enumerate}
Let us represent the estimated density function of elements in set $X$ using the KDE method as $\hat{f}_X$. Simultaneously, we denote the estimated density function of elements in the transformed set $CY+D$ by $\hat{f}_Y(C, D)$, also obtained through the KDE method. 
We will determine the optimal transformation parameters $C$ and $D$ in this non-Gaussian case by solving the following optimization problem associated with \eqref{eq:main}:
\begin{equation}\label{eq:opt}
    \inf_{C,D}\, D_{\text{KL}}\big(\hat{f}_X || \hat{f}_Y(C,D)\big).
\end{equation}
We employ the \textbf{gradient descent} method to solve Problem \eqref{eq:opt}, where the gradient of the loss function with respect to $(C, D)$ is given by:
\begin{equation}\label{eq:grad}
    \nabla\, D_{\text{KL}}\big(\hat{f}_X || \hat{f}_Y(C,D) \big) = -\int_{\mathbb{R}^d} \frac{\nabla\, \hat{f}_Y(C,D) (x)}{\hat{f}_Y (C,D) (x)} \hat{f}_X (x) \, dx.
\end{equation}
The complete procedure for the data fusion method, applicable to both Gaussian and non-Gaussian scenarios, is summarized in Algorithm~\ref{alg1} below. In this algorithm, the gradient is computed using Equation~\eqref{eq:grad}, and the stopping criterion is based on the resulting KL divergence (or gradient norm) falling below a chosen threshold.

\begin{algorithm}
\SetAlgoLined
\textbf{Input:} Feature sets $X$ and $Y$\;
\If{\textnormal{$X$ and $Y$ are Gaussian distributions}}{
    \textbf{Return} $(C^{*},D^{*})$ by evaluating Equation~\eqref{eq:gauss}\;
}
\Else{
    \textbf{Compute} $\hat{f}_X \gets \text{KDE}(X)$\;
    \textbf{Initialize} $(C, D)$ and choose a learning rate $\tau > 0$\;
    \While{\textnormal{the stopping criterion is not met}}{
        $(C, D) \gets (C, D) - \tau\, \nabla\, D_{\text{KL}}\Bigl(\hat{f}_X \,\Vert\, \hat{f}_Y(C, D)\Bigr)$\;
    }
    \textbf{Return} $(C^{*}, D^{*}) = (C, D)$.
}
\caption{Data Fusion Algorithm}\label{alg1}
\end{algorithm}

\subsection*{Classification procedure}

After the feature based fusion process, we start the classification task. The fused features' dataset is uniform in distribution and will be fed into the classification algorithm. The algorithm recognizes the inherent patterns and features of the dataset so as to accurately classify the data points into predefined categories.
For each feature, we conducted 100 experiments In each experiment, we randomly selected a subset of 1000 data points from the health dataset (labelled 1) and the disease dataset (labelled 0) to ensure a balanced representation of each category. To ensure reproducible results, the random number generator used a fixed seed for random selection.The classification effectiveness of the model is quantified by the accuracy metric, which reflects the proportion of correctly classified instances in all cases. In this way, we can assess the consistency of the performance of the classifiers in various cases and thus get a comprehensive picture of their predictive power.

\section*{Data Availability}
The in-house data used in this study are available from the corresponding author upon a reasonable request and subject to applicable ethical and legal restrictions. Due to the sensitive nature of the data, access may be granted on a case-by-case basis following institutional and regulatory approvals.

\section*{Code Availability}
The code available from the corresponding author on a reasonable request and subject to an institutional approval.

\section*{Acknowledgment}
The study is supported by the seed fund of Aerie Intelligent Technology Corp (USA).

\section*{Author Contributions}
B.S. developed the code, conducted the research, and drafted manuscript. Q.D. and Y.G. collected data and participated in experiment design. T.T. and K.L. designed the model and supervised the operations. All authors reviewed and revised the manuscript and consent to publication.

\section*{Competing Interests}
The authors declare no competing interests.

\section*{Appendix}

\begin{table}[htbp]
\centering
\textbf{Table 5 | }Summary Datasets used of In-house datasets \\ 
\begin{tabular}{|c|c|c|c|}
\hline
Dataset & Sample size&Length $\times$ dimensions & Sampling frequency \\
\hline
Health  I & 1000&10000 $\times$  12 & 500 Hz \\
\hline
Health  II & 1000&5000 $\times$ 12 & 1000 Hz \\
\hline
Disease I & 2000&5000 $\times$  12 & 1000 Hz \\
\hline
\end{tabular}
\label{tab:datasets_summary_of_In-house datasets}
\end{table}

\begin{table}[htbp]
    \centering
    \textbf{Table 6 | } Summary of Datasets used of PTB-XL datasets
    \begin{tabular}{|c|c|c|c|}
    \hline
    Dataset& Sample size& Length$\times$ dimensions&Sampling frequency\\
    \hline
     Norm I &1000&
     2500$\times$12&500 Hz\\
     \hline
     Norm I &1000&500$\times$12&100 Hz\\
     \hline
     Disease I & 1000 & 2500$\times$12&100 Hz\\
     \hline
     Disease II& 1000 &2500$\times$12&500 Hz\\
     \hline
    \end{tabular}
    \label{tab:datasets_summary_of_PTBXL_Datasets}
\end{table}

\bibliography{ref}

\end{document}